\overfullrule=0pt
\input harvmac
\def\a{{\alpha}}

\def\l{{\lambda}}

\def\b{{\beta}}

\def\g{{\gamma}}

\def\d{{\delta}}
\def\e{{\epsilon}}

\def\p{{\partial}}

\def\t{{\theta}}

\def\bar{\overline}
\def\({\left(}
\def\){\right)}

\Title{\vbox{\hbox{IFT-P.000/2005 }}}
{\vbox{
\centerline{\bf Four-Point One-Loop Amplitude Computation  }
\centerline{\bf in the Pure Spinor Formalism }}}
\bigskip\centerline{
 	Carlos R. Mafra\foot{e-mail: crmafra@ift.unesp.br}
 }
\bigskip
\centerline{\it Instituto de F\'\i sica Te\'orica, Universidade Estadual
Paulista}
\centerline{\it Rua Pamplona 145, 01405-900, S\~ao Paulo, SP, Brasil}

\vskip .3in
The massless 4-point one-loop amplitude computation in the pure spinor
formalism is shown to agree with the computation in the RNS formalism.

\vskip .3in

\Date {December 2005}


\newsec{Introduction}

In the year 2000, Berkovits proposed a new formalism for the superstring
with manifest space-time supersymmetry that can be covariantly 
quantized\ref\superp{N. Berkovits, 
{\it Super-Poincar\'e Covariant Quantization of the
Superstring}, JHEP 04 (2000) 018, hep-th/0001035.}.
Since then, the formalism has evolved to a point where multiloop 
superstring amplitudes are computed in a manifestly super-Poincar\'e
manner \ref\nathanmultiloop{N. Berkovits, {\it Multiloop Amplitudes and
Vanishing Theorems using
the Pure Spinor Formalism for the Superstring}, JHEP 0409 (2004) 047,
hep-th/0406055.} with relative easy when compared with the RNS formalism.
In the last five years 
there have been lots of consistency checks, and up to now
the pure spinor formalism has bravely survived. The last one 
of these checks was the agreement with the RNS result 
for massless 4-point two-loop 
amplitudes \ref\nathantwoloop{N. Berkovits,
{\it Super-Poincar\'e Covariant Two-Loop Superstring Amplitudes},
hep-th/0503197.}\ref\twoloop{N. Berkovits and
C. Mafra {\it Equivalence of Two-Loop Superstring Amplitudes in the Pure Spinor
and RNS Formalisms}, hep-th/0509234}\ref\dhoker{E. D'Hoker and D.H. Phong, 
{\it Two Loop Superstrings,
1. Main Formulas}, Phys. Lett. B529 (2002)
241, hep-th/0110247\semi
E. D'Hoker and D.H. Phong,
{\it Two Loop Superstrings, 2. The Chiral Measure on Moduli Space}, Nucl.
Phys. B636 (2002) 3, hep-th/0110283\semi
E. D'Hoker and D.H. Phong,
{\it Two Loop Superstrings, 3. Slice Independence and Absence
of Ambiguities}, Nucl. Phys. B636 (2002) 61, hep-th/0111016\semi
E. D'Hoker and D.H. Phong,
{\it Two Loop Superstrings, 4. The Cosmological Constant and Modular
Forms},
Nucl. Phys. B639 (2002) 129, hep-th/0111040\semi
E. D'Hoker and D.H. Phong,
{\it Two Loop Superstrings, 5. Gauge-Slice Independendence of the N-Point
Function}, hep-th/0501196\semi
E. D'Hoker and D.H. Phong,
{\it Two Loop Superstrings, 6. Non-Renormalization Theorems and the
Four-Point Function}, hep-th/0501197.}\ref\zhuold{
R. Iengo and C.-J. Zhu, {\it
Two Loop Computation of the Four-Particle
Amplitude in the Heterotic String}, Phys. Lett. B212 (1988) 313\semi
R. Iengo and C.-J. Zhu, {\it Explicit Modular Invariant Two-Loop Superstring
Amplitude Relevant for $R^4$}, JHEP 06 (1999) 011, hep-th/9905050
\semi R. Iengo, {\it
Computing the $R^4$ Term at Two Superstring Loops}, JHEP 0202 (2002) 035,
hep-th/0202058\semi
Z.-J. Zheng, J.-B. Wu and C.-J. Zhu, {\it
Two-Loop Superstrings in Hyperelliptic Language I: the Main
Results}, Phys. Lett. B559 (2003) 89, hep-th/0212191\semi
Z.-J. Zheng, J.-B. Wu and C.-J. Zhu, {\it
Two-Loop Superstrings in Hyperelliptic Language II: the Vanishing
of the Cosmological Constant and the Non-Renormalization Theorem},
Nucl. Phys. B663 (2003) 79, hep-th/0212198\semi
Z.-J. Zheng, J.-B. Wu and C.-J. Zhu, {\it
Two-Loop Superstrings in Hyperelliptic Language III: the Four-Particle
Amplitude}, Nucl. Phys. B663 (2003) 95, hep-th/0212219\semi
W.-J. Bao and C.-J. Zhu, {\it Comments on Two-Loop Four-Particle
Amplitude in Superstring Theory}, JHEP 0305 (2003) 056, hep-th/0303152.}
(see also \ref\sduality{
  E.~D'Hoker, M.~Gutperle and D.~H.~Phong,
  {\it Two-loop superstrings and S-duality},
  Nucl.\ Phys.\ B {\bf 722}, 81 (2005)
  [arXiv:hep-th/0503180].}).

The one-loop agreement has already been considered in
\ref\anguelova{L. Anguelova,
P.A. Grassi and P. Vanhove, {\it Covariant One-Loop Amplitudes in $D=11$},
Nucl. Phys. B702 (2004) 269, hep-th/0408171.},
where it was argued that the pure spinor amplitude coincides 
with the RNS result of \ref\schwarz{M.B.Green and J.H. Schwarz,
{\it Supersymmetric Dual String Theory (III): Loops and Renormalization} 
Nucl. Phys. B198 (1982) 441} for constant field-strength. 
However, we will show that there are subtleties in the computation 
at zero momentum and that the naive computation of \anguelova\ 
gives the wrong answer. In this paper we will perform this computation for non-constant
field-strength and will obtain complete agreement with the RNS computation.

\newsec{Massless 4-point one-loop amplitude in
the pure spinor formalism}

In \nathanmultiloop\ Berkovits obtained the following formula for
the massless 4-point one-loop amplitude for the type-IIB superstring,
which we rewrite in a slightly different fashion as,
\eqn\fourptoneloop{
{\cal A} = K{\bar K}\int {d^2\tau \over ({\rm Im}\tau)^2} F_c(\tau),
}
where $F_c(\tau)$ is a modular invariant function defined
by \ref\gswII{Green, M. B., Schwarz, J. H., Witten, E. {\it 
Superstring Theory: 2. Loop Amplitudes, Anomalies \& Phenomenology}.
Cambridge University Press (1987)}
$$
F_c(\tau)={1\over ({\rm Im}\tau)^3}\int d^2z_2\int d^2z_3\int d^2z_4
\prod_{i<j}G(z_i,z_j)^{k_i\cdot k_j},
$$
$G(z_i,z_j)$ is the scalar Green's function and $K$ is a kinematic
factor which reads
$$
K=\int d^{16}\t(\e T^{-1})^{((\a\b\g))}_{[\rho_1{\ldots} \rho_{11}]}
\t^{\rho_1}{\ldots}\t^{\rho_{11}}(\g_{mnpqr})_{\b\g}A_{1\a}(\t)
(W_2(\t)\g^{mnp}W^3(\t)){\cal F}^{qr}_4(\t) +{\rm perm}(234).
$$
Using the same trick of \twoloop, where
$\int d^{16}\t(\e T^{-1})^{((\a\b\g))}_{[\rho_1{\ldots} \rho_{11}]}
\t^{\rho_1}{\ldots}\t^{\rho_{11}}f_{\a\b\g}$ is expressed as the
tree-level pure spinor correlator $\langle \l^{\a}\l^{\b}\l^{\g}
(\t)^5D^5 f_{\a\b\g}\rangle$, 
$K$ can be rewritten as
\eqn\purekinematic{
K = \langle(\t)^5D^5 (\l A^1)(\l \g^{[m} W^2)(\l\g^{n]}W^3){\cal F}^4_{mn}\rangle
+ {\rm perm}(234).
}
When all external states are in the Neveu-Schwarz sector \purekinematic\
will be shown to coincide with the well-known
RNS result, i.e., $K^{\rm 4-NS} \propto t_8F^1F^2F^3F^4$,
where the $t_8$-tensor is
defined in \gswII.

\newsec{Equivalence with the RNS formalism}

Since $A_{\a}(\t)$ and $W^{\a}(\t)$ are fermionic while ${\cal F}(\t)$
is bosonic, the contributions when all external states are NS come
from terms in which an odd (even) number of covariant derivatives act
upon the fermionic (bosonic) superfields. One therefore has
\eqn\principium{
K^{\rm 4-NS}=(\t)^5\Big[
20D^3(\l A^1)(\l\g^{[m} DW^2)(\l\g^{n]} DW^3){\cal F}^4_{mn}+
}
$$
+60D(\l A^1)(\l\g^{[m} DW^2)(\l\g^{n]} DW^3)D^2{\cal F}^4_{mn}+
$$
$$
+20D(\l A^1)(\l\g^{[m} D^3W^2)(\l\g^{n]} DW^3){\cal F}^4_{mn}
+20D(\l A^1)(\l\g^{[m} DW^2)(\l\g^{n]} D^3W^3){\cal F}^4_{mn}\Big],
$$
where the spinor indices of $(\t)^5$ are contracted with the
covariant derivatives and the combinatoric factors in \principium\ come
from the different ways of splitting up these five indices.

Using the following relations,
$$
\t^{\a}D_{\a}W^{\b}=-{1\over 4}(\g^{mn}\t)^{\b}{\cal F}_{mn},
$$
$$
\t^{\a}\t^{\b}D_{\a}D_{\b}{\cal F}^{mn} = {1\over 4}k^{[m}(\t\g^{n]tu}\t){\cal F}_{tu},
$$
$$
\t^{\a_1}\t^{\a_2}\t^{\a_3}D_{\a_1}D_{\a_2}D_{\a_3}W^{\b}=
-{1\over 8}(\g^{mn}\t)^{\b}k_m(\t\g_n\g^{pq}\t){\cal F}_{pq}
$$
$$
\t^{\a_1}\t^{\a_2}\t^{\a_3}D_{\a_1}D_{\a_2}D_{\a_3}(\l A)=
{3\over 16}F_{mn}(\l\g_p\t)(\t\g^{mnp}\t),
$$
where \nathantwoloop\ $D_{\a} = \p_{\a} + {1\over 2}k_m(\g^m\t)_{\a}$ and 
$$
\l^{\a}A_{\a}(\t)={1\over 2}e_m(\l\g^m\t)-{1\over 3}(\xi\g_m\t)(\l\g^m\t)
-{1\over 32}F_{mn}(\l\g_p\t)(\t\g^{mnp}\t) + \ldots
$$
equation \principium\ becomes,
\eqn\computatio{
K_1^{NS} = +{15\over 64}F^1_{mn}F^2_{pq}F^3_{rs}F^4_{tu}
\langle
(\l\g^{[t}\g^{pq}\t)(\l\g^{u]}\g^{rs}\t)(\l\g_a\t)(\t\g^{mna}\t)
\rangle +
}
$$
+{15\over 16}(k^4_me^1_n)F^2_{pq}F^3_{rs}F^4_{tu}\langle(\l\g^{[m}\g^{pq}\t)
(\l\g^{a]}\g^{rs}\t)(\l\g^n\t)(\t\g_a\g^{tu}\t)\rangle +
$$
$$
+{5\over 16}(k^2_me^1_n)F^2_{pq}F^3_{rs}F^4_{tu}\langle(\l\g^{[t}\g^{ma}\t)
(\l\g^{u]}\g^{rs}\t)(\l\g^n\t)(\t\g_a\g^{pq}\t)\rangle +
$$
$$
+{5\over 16}(k^3_me^1_n)F^2_{pq}F^3_{rs}F^4_{tu}\langle(\l\g^{[t}\g^{pq}\t)
(\l\g^{u]}\g^{ma}\t)(\l\g^n\t)(\t\g_a\g^{rs}\t)\rangle.
$$
In \anguelova\ the authors ignored the last three lines of \computatio\ by
considering a constant field-strength and have reported to obtain the correct
RNS result. However, we will show that this does not happen. Agreement
with the RNS formalism is only obtained after summing up all contributions
in \computatio, and the inability to get the correct result supposing $F_{mn}$ 
constant may be related to subtleties in amplitude computations
at zero momentum, as will be commented in the last section.

Using the identity $\g^m\g^{np} = \g^{mnp} + \eta^{m[n}\g^{p]}$ one can check
that three types of correlation functions\foot{In version 3 of \anguelova, 
equation (3.3)
was not correctly obtained since all deltas in the right
hand side were ignored. Their identity for (3.4) was also wrong. After being
informed of these facts, Pierre Vanhove has independently obtained a much 
simpler
way to obtain the coefficients $A$ and $B$ than the one presented 
here \ref\vanhove{Pierre Vanhove, private
communication.}.
}
will be needed to evaluate 
\computatio\
\eqn\primum{
\langle(\l\g_{t}\t)(\l\g^{mnp}\t)(\l\g^{qrs}\t)(\t\g_{ijk}\t)\rangle=
C\e^{ijkmnpqrst}+
}
$$
+A\Big[ 
	 \d^{[m}_t\d^n_{[i}\eta^{p][q}\d^r_j\d^{s]}_{k]}
	-\d^{[q}_t\d^r_{[i}\eta^{s][m}\d^n_j\d^{p]}_{k]}
\Big]
+B\Big[ 
	 \eta_{t[i}\eta^{v[q} \d^r_j\eta^{s][m}\d^n_{k]}\d^{p]}_v 
	-\eta_{t[i}\eta^{v[m} \d^n_j\eta^{p][q}\d^r_{k]}\d^{s]}_v
\Big]
$$
\eqn\secundum{
\langle(\l\g^{mnp}\t)(\l\g_{q}\t)(\l\g_{t}\t)(\t\g_{ijk}\t)\rangle=
       {1\over 70}\d^{[m}_{[q}\eta_{t][i}\d^{n}_j\d^{p]}_{k]} 
}
\eqn\tertium{
\langle(\l\g^{m}\t)(\l\g^{n}\t)(\l\g^{p}\t)(\t\g_{ijk}\t)\rangle={1\over
120}\d^{mnp}_{ijk},
}
where $A=-2B={1\over 140}$, $C={1\over 8400}$, as will be shown in the
sequence. Furthermore,
it is not difficult to justify \secundum\ and \tertium\ by noting that they 
are the only
possible linear combinations of $\eta_{mn}$ tensors that have the 
appropriate symmetries and are compatible with the properties of the 
pure spinor $\l^{\a}$. Moreover, they are normalized such that
\eqn\norm{
\langle (\l\g_m\t)(\l\g_n\t)(\l\g_p\t)(\t\g^{mnp}\t)\rangle = 1.
}
The following identity
$$
(\l\g^{mnp}\t)(\l\g^{qrs}\t) = -{1\over 32\cdot 5!}(\l\g^{abcde}\l)
(\t\g^{mnp}\g_{abcde}\g^{qrs}\t)
$$
\eqn\ftensor{
\equiv -{1\over 32\cdot 5!}(\l\g^{abcde}\l)(\t\g^{tuv}\t)f^{mnpqrs}_{abcdetuv},
}
will allow one to determine both coefficients $A$ and $B$. From \ftensor\
it follows that
\eqn\dificilis{
\langle(\l\g_{t}\t)(\l\g^{mnp}\t)(\l\g^{qrs}\t)(\t\g_{ijk}\t)\rangle=
-{1\over 3840}\langle(\l\g^{abcde}\l)(\l\g_t\t)(\t\g_{ijk}\t)(\t\g^{uvx}\t)\rangle
f^{mnpqrs}_{abcdeuvx},
}
where the correlation function in the right hand side of \dificilis\
has already been 
determined, up to terms involving Levi-Civita's epsilons, to be \twoloop
\eqn\F{
   \langle(\l\g^{mnpqr}\l)(\l\g^{u}\t)(\t\g_{fgh}\t)
   (\t\g_{jkl}\t)\rangle =
}
$$
= -{4\over 35}\Big[
   \d^{[m}_{[j} \d^n_k \d^p_{l]}\d^q_{[f}\d^{r]}_g \d^u_{h]}
   +\d^{[m}_{[f} \d^n_g \d^p_{h]}\d^q_{[j}\d^{r]}_k \d^u_{l]}
   -{1\over 2}\d^{[m}_{[j}\d^n_k \eta_{l][f}\d^p_g\d^q_{h]}\eta^{r]u}
   -{1\over 2}\d^{[m}_{[f}\d^n_g \eta_{h][j}\d^p_k\d^q_{l]}\eta^{r]u}
\Big].
$$

In \twoloop\ it was argued that all terms containing
Levi-Civita's epsilons in the correlation function \F\ would not 
contribute to the two-loop amplitude under consideration and were 
safely ignored. However, in the present application of \F\ these
epsilon-terms will contribute non-epsilon terms to the left hand
side of \dificilis\ when they are contracted with epsilons contained
in $f^{mnpqrs}_{abcdetuv}$. One therefore needs to determine them, which
can be easily done by considering the self-duality condition
\eqn\selfdual{
\g^{mnpqr}_{\a\b}={1\over 120}\e^{mnpqrstuvx}\(\g_{stuvx}\)_{\a\b},
}
because it will relate non-epsilon with epsilon terms in \F. 
Using \selfdual\ one can obtain the complete correlation 
function \F\ that, when written out explicitly, is given by
\eqn\omnia{
\langle\(\l\g^{mnpqr}\l\)\(\l\g^u\t\)\(\t\g_{fgh}\t\)\(\t\g_{jkl}\t\)\rangle = 
}
$$
+{4\over 105}\Big[
-\d^{l}_{u}\d^{fghjk}_{mnpqr} 
+\d^{k}_{u}\d^{fghjl}_{mnpqr} 
-\d^{j}_{u}\d^{fghkl}_{mnpqr}
-\d^{h}_{u}\d^{fgjkl}_{mnpqr}
+\d^{g}_{u}\d^{fhjkl}_{mnpqr}
$$
$$
-\d^{f}_{u}\d^{ghjkl}_{mnpqr}
+{1\over 3}\d^{h}_{l}\d^{fgjku}_{mnpqr} 
-{1\over 3}\d^{h}_{k}\d^{fgjlu}_{mnpqr} 
+{1\over 3}\d^{h}_{j}\d^{fgklu}_{mnpqr}
-{1\over 3}\d^{g}_{l}\d^{fhjku}_{mnpqr}
$$
$$
+{1\over 3}\d^{g}_{k}\d^{fhjlu}_{mnpqr}
-{1\over 3}\d^{g}_{j}\d^{fhklu}_{mnpqr}
+{1\over 3}\d^{f}_{l}\d^{ghjku}_{mnpqr} 
-{1\over 3}\d^{f}_{k}\d^{ghjlu}_{mnpqr} 
+{1\over 3}\d^{f}_{j}\d^{ghklu}_{mnpqr}
\Big]
$$
$$
+ {1\over 3150}\Big[
+\d^{l}_{u}\e^{fghjkmnpqr} 
-\d^{k}_{u}\e^{fghjlmnpqr} 
+\d^{j}_{u}\e^{fghklmnpqr}
+\d^{h}_{u}\e^{fgjklmnpqr}
-\d^{g}_{u}\e^{fhjklmnpqr}
$$
$$
+\d^{f}_{u}\e^{ghjklmnpqr}
+ {1\over 3}\d^{h}_{l}\e^{fgjkmnpqru} 
-{1\over 3}\d^{h}_{k}\e^{fgjlmnpqru} 
+{1\over 3}\d^{h}_{j}\e^{fgklmnpqru}
- {1\over 3}\d^{g}_{l}\e^{fhjkmnpqru}
$$
$$
-{1\over 3}\d^{g}_{k}\e^{fhjlmnpqru} 
+{1\over 3}\d^{g}_{j}\e^{fhklmnpqru}
+{1\over 3}\d^{f}_{l}\e^{ghjkmnpqru} 
-{1\over 3}\d^{f}_{k}\e^{ghjlmnpqru} 
+{1\over 3}\d^{f}_{j}\e^{ghklmnpqru}
\Big].
$$

After finding the above identity one must obtain the explicit form
of the f-tensor \ftensor, which is straightforward in principle,
but tedious in practice. This task was done with the Mathematica 
package GAMMA \ref\GAM{U. Gran,
{\it GAMMA: A Mathematica Package for Performing Gamma-Matrix Algebra
and Fierz Transformations in Arbitrary Dimensions}, hep-th/0105086.},
along with some custom-made functions to handle Levi-Civita's
epsilons and duality relations for the gamma matrices. In particular,
the following identities were used,
$$
\(\g^{m_1m_2m_3m_4m_5m_6}\)^{\,\,\b}_{\a}=+{1\over 4!}
\e^{m_1m_2m_3m_4m_5m_6n_1n_2n_3n_4}\(\g_{n_1n_2n_3n_4}\)^{\,\,\b}_{\a}
$$
$$
\(\g^{m_1m_2m_3m_4m_5m_6m_7}\)_{\a\b}=-{1\over 3!}
\e^{m_1m_2m_3m_4m_5m_6m_7n_1n_2n_3}\(\g_{n_1n_2n_3}\)_{\a\b}
$$
$$
\(\g^{m_1m_2m_3m_4m_5m_6m_7m_8}\)^{\,\,\b}_{\a}=-{1\over 2!}
\e^{m_1m_2m_3m_4m_5m_6m_7m_8n_1n_2}\(\g_{n_1n_2}\)^{\,\,\b}_{\a}.
$$

After determining the f-tensor one can obtain the correlation function
\primum\ using equation \dificilis.  The coefficients of \primum\ are then
found to be $A=-2B={1\over 140}$, $C={1\over 8400}$. With these coefficients
one can check that the following consistency condition between
\primum\ and \secundum\ is indeed satisfied,
$$
\langle(\l\g_{t}\t)(\l\g^{tnp}\t)(\l\g^{qrs}\t)(\t\g_{ijk}\t)\rangle=
2\langle(\l\g^{qrs}\t)(\l\g^p\t)(\l\g^n\t)(\t\g_{ijk}\t)\rangle.
$$

Using the identities \primum,\secundum\ and \tertium\ the kinematic
factor \computatio\ can be straightforwardly computed. 
After summing over the permutations, 
using momentum conservation, $(k^R\cdot e^R)=0$
and expressing everything in terms of the Mandelstam 
variables $u=-2(k^1\cdot k^3)$ and $t=-2(k^2\cdot k^3)$ only, 
the first line
of \computatio\ gives the following result:
\eqn\wrong{
-{1\over 56} (k^2\cdot e^3)(k^2\cdot e^4)(k^3\cdot e^2)(k^4\cdot e^1) 
-{1\over 56} (k^2\cdot e^3)(k^3\cdot e^2)(k^3\cdot e^4)(k^4\cdot e^1)
}
$$
-{1\over 56} (k^2\cdot e^3)(k^2\cdot e^4)(k^3\cdot e^1)(k^4\cdot e^2)
+{1\over 56} (k^3\cdot e^1)(k^3\cdot e^2)(k^3\cdot e^4)(k^4\cdot e^3)
$$
$$
+{1\over 56} (k^3\cdot e^2)(k^3\cdot e^4)(k^4\cdot e^1)(k^4\cdot e^3)
-{1\over 56} (k^2\cdot e^4)(k^3\cdot e^1)(k^4\cdot e^2)(k^4\cdot e^3)
$$
$$
+{1\over 56} (k^3\cdot e^1)(k^3\cdot e^4)(k^4\cdot e^2)(k^4\cdot e^3) 
+{1\over 56} (k^3\cdot e^4)(k^4\cdot e^1)(k^4\cdot e^2)(k^4\cdot e^3)
-{11\over 168} (k^2\cdot e^3)(k^2\cdot e^4)(e^1\cdot e^2)t
$$
$$
-{11\over 168} (k^2\cdot e^4)(k^4\cdot e^3)(e^1\cdot e^2)t
-{1\over 112} (k^3\cdot e^4)(k^4\cdot e^3)(e^1\cdot e^2)t 
+{11\over 168} (k^2\cdot e^4)(k^3\cdot e^2)(e^1\cdot e^3)t
$$
$$
-{11\over 168} (k^3\cdot e^4)(k^4\cdot e^2)(e^1\cdot e^3)t
+{1\over 112} (k^2\cdot e^3)(k^3\cdot e^2)(e^1\cdot e^4)t 
+{11\over 168} (k^2\cdot e^3)(k^4\cdot e^2)(e^1\cdot e^4)t
$$
$$
+{11\over 168} (k^4\cdot e^2)(k^4\cdot e^3)(e^1\cdot e^4)t 
-{11\over 168} (k^2\cdot e^4)(k^3\cdot e^1)(e^2\cdot e^3)t 
-{1\over 112} (k^2\cdot e^4)(k^4\cdot e^1)(e^2\cdot e^3)t
$$
$$
-{1\over 112} (k^3\cdot e^4)(k^4\cdot e^1)(e^2\cdot e^3)t 
-{11\over 168} (k^2\cdot e^3)(k^4\cdot e^1)(e^2\cdot e^4)t 
-{11\over 168} (k^3\cdot e^1)(k^4\cdot e^3)(e^2\cdot e^4)t
$$
$$
-{11\over 168} (k^4\cdot e^1)(k^4\cdot e^3)(e^2\cdot e^4)t 
-{1\over 112} (k^3\cdot e^1)(k^3\cdot e^2)(e^3\cdot e^4)t 
-{1\over 112} (k^3\cdot e^2)(k^4\cdot e^1)(e^3\cdot e^4)t
$$
$$
+{19\over 336} (k^3\cdot e^1)(k^4\cdot e^2)(e^3\cdot e^4)t 
-{1\over 112} (k^4\cdot e^1)(k^4\cdot e^2)(e^3\cdot e^4)t 
+{1\over 224} (e^1\cdot e^4)(e^2\cdot e^3)t^2
$$
$$
+{11\over 336} (e^1\cdot e^3)(e^2\cdot e^4)t^2 
+{1\over 224} (e^1\cdot e^2)(e^3\cdot e^4)t^2
-{11\over 168} (k^2\cdot e^3)(k^2\cdot e^4)(e^1\cdot e^2)u
$$
$$
-{11\over 168} (k^2\cdot e^3)(k^3\cdot e^4)(e^1\cdot e^2)u 
-{1\over 112} (k^3\cdot e^4)(k^4\cdot e^3)(e^1\cdot e^2)u 
+{11\over 168} (k^2\cdot e^4)(k^3\cdot e^2)(e^1\cdot e^3)u
$$
$$
+{11\over 168} (k^3\cdot e^2)(k^3\cdot e^4)(e^1\cdot e^3)u 
+{1\over 112} (k^2\cdot e^4)(k^4\cdot e^2)(e^1\cdot e^3)u 
+{11\over 168} (k^2\cdot e^3)(k^4\cdot e^2)(e^1\cdot e^4)u
$$
$$
-{11\over 168} (k^3\cdot e^2)(k^4\cdot e^3)(e^1\cdot e^4)u 
-{11\over 168} (k^2\cdot e^4)(k^3\cdot e^1)(e^2\cdot e^3)u 
-{11\over 168} (k^3\cdot e^1)(k^3\cdot e^4)(e^2\cdot e^3)u
$$
$$
-{11\over 168} (k^3\cdot e^4)(k^4\cdot e^1)(e^2\cdot e^3)u 
-{1\over 112} (k^2\cdot e^3)(k^3\cdot e^1)(e^2\cdot e^4)u 
-{11\over 168} (k^2\cdot e^3)(k^4\cdot e^1)(e^2\cdot e^4)u
$$
$$
-{1\over 112} (k^3\cdot e^1)(k^4\cdot e^3)(e^2\cdot e^4)u 
-{1\over 112} (k^3\cdot e^1)(k^3\cdot e^2)(e^3\cdot e^4)u 
+{19 \over 336} (k^3\cdot e^2)(k^4\cdot e^1)(e^3\cdot e^4)u
$$
$$
-{1\over 112} (k^3\cdot e^1)(k^4\cdot e^2)(e^3\cdot e^4)u 
-{1\over 112} (k^4\cdot e^1)(k^4\cdot e^2)(e^3\cdot e^4)u 
+{11\over 336} (e^1\cdot e^4)(e^2\cdot e^3)tu
$$
$$
+{11\over 336} (e^1\cdot e^3)(e^2\cdot e^4)tu 
-{1\over 42} (e^1\cdot e^2)(e^3\cdot e^4)tu 
+{11\over 336} (e^1\cdot e^4)(e^2\cdot e^3)u^2
$$
$$
+{1\over 224} (e^1\cdot e^3)(e^2\cdot e^4)u^2
+{1\over 224} (e^1\cdot e^2)(e^3\cdot e^4)u^2
$$
which is clearly seen not to be proportional to $t_8F^1F^2F^3F^4$, as incorrectly
claimed in \anguelova.
Repeating the same procedure for the second line of \computatio\ one obtains:
\eqn\wrongtwo{
+{1 \over 56 }(k^2\cdot e^3)(k^2\cdot e^4)(k^3\cdot e^2)(k^4\cdot e^1)
+ {1 \over 56 } (k^2\cdot e^3)(k^3\cdot e^2)(k^3\cdot e^4)(k^4\cdot e^1)
}
$$
+ {1 \over 56 } (k^2\cdot e^3)(k^2\cdot e^4)(k^3\cdot e^1)(k^4\cdot e^2)
- {1 \over 56 } (k^3\cdot e^1)(k^3\cdot e^2)(k^3\cdot e^4)(k^4\cdot e^3)
$$
$$
- {1 \over 56 } (k^3\cdot e^2)(k^3\cdot e^4)(k^4\cdot e^1)(k^4\cdot e^3) 
+ {1 \over 56 } (k^2\cdot e^4)(k^3\cdot e^1)(k^4\cdot e^2)(k^4\cdot e^3)
$$
$$
- {1 \over 56 } (k^3\cdot e^1)(k^3\cdot e^4)(k^4\cdot e^2)(k^4\cdot e^3) 
- {1 \over 56 } (k^3\cdot e^4)(k^4\cdot e^1)(k^4\cdot e^2)(k^4\cdot e^3) 
- {19 \over 672 } (k^2\cdot e^3)(k^2\cdot e^4)(e^1\cdot e^2)t
$$
$$
- {19 \over 672 } (k^2\cdot e^4)(k^4\cdot e^3)(e^1\cdot e^2)t 
+ {1 \over  112 } (k^3\cdot e^4)(k^4\cdot e^3)(e^1\cdot e^2)t 
+ {19 \over 672 } (k^2\cdot e^4)(k^3\cdot e^2)(e^1\cdot e^3)t
$$
$$
- {19 \over 672 } (k^3\cdot e^4)(k^4\cdot e^2)(e^1\cdot e^3)t 
- {1 \over 112 } (k^2\cdot e^3)(k^3\cdot e^2)(e^1\cdot e^4)t 
+ {19 \over 672 } (k^2\cdot e^3)(k^4\cdot e^2)(e^1\cdot e^4)t
$$
$$
+ {19 \over 672 } (k^4\cdot e^2)(k^4\cdot e^3)(e^1\cdot e^4)t 
- {19 \over 672 } (k^2\cdot e^4)(k^3\cdot e^1)(e^2\cdot e^3)t 
+ {1 \over 112 } (k^2\cdot e^4)(k^4\cdot e^1)(e^2\cdot e^3)t
$$
$$
+ {1 \over 112 } (k^3\cdot e^4)(k^4\cdot e^1)(e^2\cdot e^3)t 
- {19 \over 672 } (k^2\cdot e^3)(k^4\cdot e^1)(e^2\cdot e^4)t 
- {19 \over 672 } (k^3\cdot e^1)(k^4\cdot e^3)(e^2\cdot e^4)t
$$
$$
- {19 \over 672 } (k^4\cdot e^1)(k^4\cdot e^3)(e^2\cdot e^4)t 
+ {1 \over 112 } (k^3\cdot e^1)(k^3\cdot e^2)(e^3\cdot e^4)t 
+ {1 \over 112 } (k^3\cdot e^2)(k^4\cdot e^1)(e^3\cdot e^4)t
$$
$$
+ {25 \over 672 } (k^3\cdot e^1)(k^4\cdot e^2)(e^3\cdot e^4)t 
+ {1 \over 112 } (k^4\cdot e^1)(k^4\cdot e^2)(e^3\cdot e^4)t
- {1 \over 224 } (e^1\cdot e^4)(e^2\cdot e^3)t^2
$$
$$
+ {19 \over 1344 } (e^1\cdot e^3)(e^2\cdot e^4)t^2
- {1 \over 224 } (e^1\cdot e^2)(e^3\cdot e^4)t^2
- {19 \over 672 } (k^2\cdot e^3)(k^2\cdot e^4)(e^1\cdot e^2)u
$$
$$
- {19 \over 672 } (k^2\cdot e^3)(k^3\cdot e^4)(e^1\cdot e^2)u 
+ {1 \over  112 } (k^3\cdot e^4)(k^4\cdot e^3)(e^1\cdot e^2)u 
+ {19 \over 672 } (k^2\cdot e^4)(k^3\cdot e^2)(e^1\cdot e^3)u
$$
$$
+ {19 \over 672 } (k^3\cdot e^2)(k^3\cdot e^4)(e^1\cdot e^3)u 
- {1 \over  112 } (k^2\cdot e^4)(k^4\cdot e^2)(e^1\cdot e^3)u 
+ {19 \over 672 } (k^2\cdot e^3)(k^4\cdot e^2)(e^1\cdot e^4)u
$$
$$
- {19 \over 672 } (k^3\cdot e^2)(k^4\cdot e^3)(e^1\cdot e^4)u 
- {19 \over 672 } (k^2\cdot e^4)(k^3\cdot e^1)(e^2\cdot e^3)u 
- {19 \over 672 } (k^3\cdot e^1)(k^3\cdot e^4)(e^2\cdot e^3)u
$$
$$
- {19 \over 672 } (k^3\cdot e^4)(k^4\cdot e^1)(e^2\cdot e^3)u 
+ {1 \over  112 } (k^2\cdot e^3)(k^3\cdot e^1)(e^2\cdot e^4)u 
- {19 \over 672 } (k^2\cdot e^3)(k^4\cdot e^1)(e^2\cdot e^4)u
$$
$$
+ {1 \over  112 } (k^3\cdot e^1)(k^4\cdot e^3)(e^2\cdot e^4)u 
+ {1 \over  112 } (k^3\cdot e^1)(k^3\cdot e^2)(e^3\cdot e^4)u 
+ {25 \over 672 } (k^3\cdot e^2)(k^4\cdot e^1)(e^3\cdot e^4)u
$$
$$
+ {1 \over  112 } (k^3\cdot e^1)(k^4\cdot e^2)(e^3\cdot e^4)u 
+ {1 \over  112 } (k^4\cdot e^1)(k^4\cdot e^2)(e^3\cdot e^4)u 
+ {19 \over 1344 } (e^1\cdot e^4)(e^2\cdot e^3)tu
$$
$$
+ {19 \over 1344 } (e^1\cdot e^3)(e^2\cdot e^4)tu 
- {31 \over 1344 } (e^1\cdot e^2)(e^3\cdot e^4)tu 
+ {19 \over 1344 } (e^1\cdot e^4)(e^2\cdot e^3)u^2
$$
$$
- {1 \over 224 } (e^1\cdot e^3)(e^2\cdot e^4)u^2 
- {1 \over 224 } (e^1\cdot e^2)(e^3\cdot e^4)u^2
$$
which is again not proportional to $t_8F^1F^2F^3F^4$. Note, however,
that the sum of \wrong\ and \wrongtwo\ is:
$$
{3\over 32}\Big[
-  (k^2\cdot e^3)(k^2\cdot e^4)(e^1\cdot e^2)t 
-  (k^2\cdot e^4)(k^4\cdot e^3)(e^1\cdot e^2)t 
+ (k^2\cdot e^4)(k^3\cdot e^2)(e^1\cdot e^3)t
$$
$$
- (k^3\cdot e^4)(k^4\cdot e^2)(e^1\cdot e^3)t 
+ (k^2\cdot e^3)(k^4\cdot e^2)(e^1\cdot e^4)t 
+ (k^4\cdot e^2)(k^4\cdot e^3)(e^1\cdot e^4)t
$$
$$
- (k^2\cdot e^4)(k^3\cdot e^1)(e^2\cdot e^3)t 
- (k^2\cdot e^3)(k^4\cdot e^1)(e^2\cdot e^4)t 
- (k^3\cdot e^1)(k^4\cdot e^3)(e^2\cdot e^4)t
$$
$$
- (k^4\cdot e^1)(k^4\cdot e^3)(e^2\cdot e^4)t 
+ (k^3\cdot e^1)(k^4\cdot e^2)(e^3\cdot e^4)t 
+ (k^3\cdot e^2)(k^4\cdot e^1)(e^3\cdot e^4)u
$$
$$
- (k^2\cdot e^3)(k^2\cdot e^4)(e^1\cdot e^2)u 
- (k^2\cdot e^3)(k^3\cdot e^4)(e^1\cdot e^2)u 
+ (k^2\cdot e^4)(k^3\cdot e^2)(e^1\cdot e^3)u
$$
$$
+ (k^3\cdot e^2)(k^3\cdot e^4)(e^1\cdot e^3)u 
+ (k^2\cdot e^3)(k^4\cdot e^2)(e^1\cdot e^4)u 
- (k^3\cdot e^2)(k^4\cdot e^3)(e^1\cdot e^4)u
$$
$$
- (k^2\cdot e^4)(k^3\cdot e^1)(e^2\cdot e^3)u 
- (k^3\cdot e^1)(k^3\cdot e^4)(e^2\cdot e^3)u 
- (k^3\cdot e^4)(k^4\cdot e^1)(e^2\cdot e^3)u
$$
$$
- (k^2\cdot e^3)(k^4\cdot e^1)(e^2\cdot e^4)u
+ {1\over 2}(e^1\cdot e^3)(e^2\cdot e^4)t^2
+ {1\over 2}(e^1\cdot e^4)(e^2\cdot e^3)u^2
$$
$$
+ {1\over 2}(e^1\cdot e^4)(e^2\cdot e^3)tu
+ {1\over 2}(e^1\cdot e^3)(e^2\cdot e^4)tu
- {1\over 2}(e^1\cdot e^2)(e^3\cdot e^4)tu
\Big]
$$
which can be checked to be proportional to $t_8F^1F^2F^3F^4$.

Repeating these same steps one can check that the last 2 lines of 
\computatio, after summing over the permutations, will also 
independently add up to a combination proportional to the RNS result.
The equivalence with the RNS formalism, after summing up {\it all}
contributions in \computatio, is then established.

\newsec{On the result}

A few comments regarding the calculations done here can be made.
There should be no doubt that to obtain equivalence with the RNS
result, all terms in \computatio\ must be considered. 
If the assumption of a constant field-strength is made and 
only the first term in \computatio\ is computed, one will obtain
the wrong answer \wrong. 

There are some possible explanations for this odd-looking fact, which
certainly deserve further investigation. 
The discussion in \ref\contact{M. B. Green, N. Seiberg 
{\it Contact Interactions in Superstring Theory}, Nucl. Phys. B299
(1988) 559} emphasizes the subtleties related to 
amplitude computations at zero momentum and explains that naive 
computations give
incorrect results because of contact terms, and that the correct 
procedure is to analyticaly continue computations at non-zero
momentum.

There may be another possible subtlety that was overlooked in 
the computation
of \anguelova. The polarization vector of a constant field-strength
is given by $e_m(X)=F_{mn}X^n$, so one is explicitly introducing 
the center-of-mass  mode of $X^m$ in the vertex 
operator. However, as explained in \ref\belo{A. Astashkevich, A. Belopolsky. 
{\it String center of mass operator and its effect on BRST cohomology}, 
Commun. Math. Phys. 186 (1997) 109, hep-th/9511111.}, the BRST cohomology
is modified if one allows vertex operators and gauge parameters involving
the center-of-mass mode of $X^m$. For example, 
one of the central tenets of the pure spinor formalism is
that the cohomology of the BRST operator 
$Q_{BRST}=\oint \l^{\a}d_{\a}$ at ghost number three is given by 
$(\l^3\t^5)\equiv (\l\g^m\t)(\l\g^n\t)(\l\g^p\t)(\t\g_{mnp}\t)$.
However, when the center-of-mass mode of $X^m$ is present, the 
ghost number three cohomology of $Q_{BRST}$ turns out to be trivial
because \ref\disse{Nathan Berkovits, private communication.},
\eqn\explanatio{
(\l^3\t^5) = Q_{BRST}\Big[X^m(\l\g^n\t)(\l\g^p\t)(\t\g_{mnp}\t)\Big].
}

So to avoid the above subtleties, in this paper the kinematic factor 
of the
massless 4-point one-loop amplitude in the pure spinor formalism
was computed with a non-constant field-strength. Equivalence
with the RNS formalism computation of \schwarz\ was correctly obtained
when all external particles are in the Neveu-Schwarz sector.

\vskip 15pt
{\bf Acknowledgements:} 
I would like to thank Nathan Berkovits for patiently helping me
on crucial stages of this work and Pierre Vanhove for discussions
concerning \anguelova.
I would also like to acknowledge
FAPESP grant 04/13290-8 for financial support.

\listrefs
\end